\documentclass{ws-ijqi-a}
\nonstopmode
\usepackage{amssymb}
\usepackage{enumerate}
\begin{document}
\markboth{Mladen Pavi{\v c}i{\'c}}
{Entanglement and Superdense Coding}
\title{Entanglement and Superdense Coding with Linear Optics}

\author{MLADEN PAVI\v CI\'C}

\address{Chair of Physics, Faculty of Civil Engineering,
University of Zagreb, Croatia\\
pavicic@grad.hr}

\maketitle

\begin{abstract}
We discuss a scheme for a full superdense coding of entangled 
photon states employing only linear-optics elements. 
By using the mixed basis consisting of four states that are 
unambiguously distinguishable by a standard and polarizing 
beam splitters we can deterministically transfer four messages 
by manipulating just one of the two entangled photons. 
The sender achieves the determinism of the transfer either by 
giving up the control over 50\%\ of sent messages (although 
known to her) or by discarding 33\%\ of incoming photons. 
\end{abstract}


\section{Introduction}
\label{intro}

Superdense coding (SC)~\cite{bennett-wiesner-92}---sending up 
to two bits of information, i.e., four messages, by manipulating 
just one of two entangled qubits (two-state quantum systems)---is 
considered to be a protocol that launched the field of quantum 
communication~\cite{barreiro-kwiat-08}. Apart from showing 
how different quantum coding of information is from the 
classical one---which can encode only two messages in a 
two-state system---the protocol has also shown how  important 
entanglement of qubits is for their manipulation.

Such an entanglement has proven to be a genuine quantum 
effect that cannot be achieved with the help of two classical 
bit carriers because we cannot entangle classical systems. 
To use this advantage of quantum 
information transfer, it is very important to keep the 
trade-off of the increased transfer capacity balanced with the 
technology of implementing the protocol. The simplest and most
efficient implementation is the one that would use 
photons manipulated by linear-optics elements such as 
beam splitters, polarizers, and wave plates and only one 
degree of freedom---polarization. 

Since entangled qubits applied to the teleportation 
required Bell states, all subsequent attempts to implement 
SC---as another transportation 
protocol---concentrated on Bell states. The idea was to send 
four messages via four Bell states [see Eq.~(\ref{eq:bell-states})] 
and herewith achieve a $\log_24=2$ bit transfer. To this aim, 
a recognition of all four Bell states was required.  

The first linear-optics implementation has reached only 
three quarters (3 messages) of its theoretical 2 bit 
(four messages) channel capacity, i.e., $\log_23=1.585$ bits. 
This was because a recognition of two Bell states $|\Psi^+\rangle$ 
and $|\Psi^-\rangle$ was achieved while the other two $|\Phi^\pm\rangle$ 
that could not be told apart were both used to send one and the same 
message. ~\cite{mattle-zeil-96} This partial realization of the 
superdense protocol was named {\em dense coding\/}. 
In 2001 Calsamiglia and L\"utkenhaus 
proved~\cite{calsamiglia-luetkenhaus-01} that the dense
coding  was all we could achieve with 
Bell states and linear optics. 

Therefore in Ref.~\cite{pavicic-code11} we dispensed with 
the Bell state basis and introduced the {\em mixed basis\/} 
which enabled us to go around the  Calsamiglia-L\"utkenhaus 
no-go proof and carry a superdense coding with linear 
optics. 

The finding revealed that the notion of superdense coding 
was not operationally well defined, mostly because no 
particular application of this protocol in quantum computation 
and/or quantum communication has been found so far. 

In this paper we therefore consider three possible operational 
definitions and implementation of the superdense coding. 

\section{Mixed Basis and Entanglement}
\label{sec:mixed-b}

We define a mixed basis as a basis which consists
of the following two Bell states
\begin{eqnarray}
|\chi^{1,2}\rangle=|\Psi^\pm\rangle=(|H\rangle_1|V\rangle_2\pm|V\rangle_1H\rangle_2)/\sqrt{2}
\label{eq:bell-states} 
\end{eqnarray}
and the following two computational basis states
\begin{eqnarray}
|\chi^3\rangle=|H\rangle_1|H\rangle_2, \qquad |\chi^4\rangle=|V\rangle_1|V\rangle_2,
\label{eq:com-states} 
\end{eqnarray}
where $H\ (V)$ represents horizontal (vertical) photon polarization.  
We shall not use the other two Bell states
$|\Phi^\pm\rangle=(|\chi^3\rangle\pm |\chi^4\rangle)/\sqrt{2}$.
Both Bell and computational bases can be expressed by means of the
mixed basis. 

Let us first see why we cannot use only the computational 
basis, then why we cannot use only the Bell basis, and in 
the end why we {\em can} use the mixed basis. We consider 
photons being sent to a beam splitter after which we try to 
split them with the help of polarizing beam splitters (PBS) 
and then detect them by means of detectors with photon 
number resolution. 

When we send two parallelly polarized photons
to a beam splitter from its opposite sides
they will always emerge from the same side, bunched
together and showing the so called Hong-Ou-Mandel
interference dip.~\cite[Sec.~3.2]{ou-book07} It has been
calculated that both bunched photons keep the polarization
direction they had before they entered the beam
splitter.~\cite{p-pra94,p-s94,pavicic-book-05}
So we can discriminate  $|\chi^3\rangle$ and $|\chi^4\rangle$
from each other and from $|\chi^{1,2}\rangle$ with  photon
number resolution detectors or up to an arbitrary precision
with single photon detectors. If we sent perpendicularly 
polarized photons---the other
two states of the computational basis---to a beam
splitter, they would either bunch together (50\%) or
emerge from the opposite sides  of the beam slitter
(50\%).~\cite{p-pra94}
The two photons that are split are correlated but unpolarized.
Therefore we cannot distinguish between $|HV\rangle$ and
$|VH\rangle$ in 50\%\ of the events and we again end up with the
channel capacity $\log_2 3$ as for the Bell states.

On the other hand, in the Bell basis we can discriminate between
$|\Psi^+\rangle$ and $|\Psi^-\rangle$ but not between
$|\Phi^+\rangle$ and $|\Phi^-\rangle$.~\cite[Sec.~4.1]{ou-book07}
At a polarization preserving (metallic) BS, $|\Psi^-\rangle$ photons
split and $|\Psi^+\rangle$ photons bunch together but have different
polarization so that we can split them at polarizing beam splitters (PBSs)
behind BS. $|\Phi^\pm\rangle$ photons also bunch together
but being entangled (unpolarized but correlated in polarization)
both photons from a pair project to either $|H\rangle$ or
$|V\rangle$, i.e., either both go through or are both reflected from
PBSs.

So, we can unambiguously discriminate two states from the 
computational basis, $|HH\rangle,|VV\rangle$, two from the 
Bell basis, $|\Psi^\pm\rangle$,  as well as as any one of them 
from each other by means of photon number resolution detectors. 
Thus we can discriminate all four $|\chi^i\rangle$,
$i=1,\dots,4$ and now there comes the question how to 
prepare them. 

Alice gets $|\Psi^+\rangle$ photons by menas of spontaneous 
parametric down-conversion 
in a BBO crystal.~\cite{mattle-zeil-96}. To send  
$|\chi^1\rangle=|\Psi^+\rangle$ she puts nothing in the 
path of her photon. To send $|\chi^2\rangle=|\Psi^-\rangle$ she puts 
in ${\rm HWP}(0^\circ)$ (halfwave plate) in the path. It changes the
sign of the vertical polarization.  To send  $|\chi^3\rangle$ she
takes out ${\rm HWP}(0^\circ)$ and puts in ${\rm  HWP}(45^\circ)$
and a polarizer ({\em pol\/}) oriented horizontally. Her {\em pol\/}
is of a polarizing beam splitter (PBS) type: $|H\rangle$ photon 
passes through and $|V\rangle$ is reflected from it. 
${\rm  HWP}(45^\circ)$ turns $|\Psi^+\rangle$ into $|\Phi^-\rangle$ 
and {\em pol\/} projects both photons to state $|H\rangle$ in half 
of the occurrences. In the other half of the occurrences Alice's 
photon is reflected from her {\em pol} and we have both photons in 
state $|V\rangle$, i.e, the pair in state $|\chi^4\rangle$. 
Alice might detect these ``wrong'' photons with the help of 
a detectors $d$. Below we will specify what Alice can do next. 
To send $|\chi^4\rangle$, Alice is making use of a reflection from 
her PBS and then the ``wrong'' photons go through. 

We stress here that the  preparation of $|\chi^3\rangle$  and $|\chi^4\rangle$
includes physics of entangled systems because whenever Alice sends her
qubit through a polarizer oriented horizontally or vertically, the
other qubit from the entangled pair (originally in the state
$|\Phi^+\rangle$) will be immediately set into $|H\rangle$ and
$|V\rangle$ state for any subsequent measurement along $H$ or
$V$ directions, respectively.

We noticed above that ``in a way'' Alice does not have a control 
over the choice of her photon while preparing $|\chi^3\rangle$  
and $|\chi^4\rangle$ states. Her photon can go either way in her 
PBS. But she does know which way it took {\it after\/} it did so. 
And this opens a question of an operational 
definitions and implementation of the superdense coding.

\section{Operational Definitions of Superdense Coding}
\label{sec:def}

In the absence of a well defined application there can be three 
possible operational definitions and implementations of superdense 
coding.  

We start with a formal definition. 
\begin{definition}\label{def:superdense}
Superdense coding is a technique used in quantum information theory to
send two bits of classical information using only one qubit, with the aid of entanglement.
\end{definition}

To make this definition more operational we restate it following 
Ref.~\cite{superdense-wolfram-web}.
\begin{definition}\label{def:superdense-wolfram}
In superdense coding, a sender (Alice) can send a message consisting
of two classical bits using one quantum bit (qubit) to the receiver
(Bob). The input to the circuit is one of a pair of qubits entangled
in the Bell basis state. The other qubit from the pair is sent
unchanged to Bob. After processing the former qubit in one of four 
ways, it is sent to Bob, who measures the two qubits, yielding two 
classical bits. The result is that Bob receives two classical bits 
which match those that Alice sent by manipulating just her qubit.
\end{definition}

Even this definition is not operational enough because several 
elements remained unspecified: 
\begin{enumerate}[(i)]\item the owner of the pair can be Alice, Bob, or 
Anna;
\item Alice might be required to send each photon she receives 
from the source to Bob or might not be required to do so;
\item Alice might be required to have a control over sent messages
or not.
\end{enumerate}

We shall consider all the aforementioned options.

In the original version of their superdense coding, Bennett and 
Wiesner~\cite{bennett-wiesner-92} assume that Bob is the 
proprietor of entangled pairs and that he sends Alice one qubit 
from each of his pairs. She manipulates her qubits and sends 
it back to Bob. Bob expects of Alice to return him each qubit 
he sent her. In our version she might do that [option (b)] or 
might not do so [options (a) and (c)]. 

In the first coding experiment  \cite{mattle-zeil-96}, Alice 
[Bob in the cited reference]  owns the  entangled pairs, 
manipulates one of the qubits, and then sends  both qubits 
of Bob [Alice in the cited reference]. In our version Alice 
might [option (c)] or might not [options (a) and (b)] 
own the pairs. 
 
Three possible scenarios that operationalize the options are:
\begin{enumerate}[(a)]
\item Alice is assumed to send Bob a comprehensible 
message by means of four elementary messages $|\chi^i\rangle$, 
$i=1,\dots,4$. Bob is the owner of the source; he sends one photon to
Alice and keeps one for himself. (A) She manipulates her qubits 
and sends to Bob only those ones over which she can have a 
control; she discards those over which she cannot have a 
control. Anna might also be the owner of the source. 
She sends one qubit to Bob and one to Alice and they proceed as 
from point (A) above;
\item Alice is assumed to send Bob an intelligible but  
not necessarily a comprehensible message by means of 
four elementary messages $|\chi^i\rangle$, $i=1,\dots,4$.
Anna owns the BBO crystal and sends one qubit to Bob 
and one to Alice. (B) Alice sends either original or cloned qubits to 
Bob; she does not have a control over 50\%\ of her 
messages (assuming they are evenly distributed) but she 
does have records of all the messages she sent. Alternatively, 
Bob can own the source and send one photon to
Alice and keeps one for himself. Then they proceed as from 
point (B) above;
\item Alice owns the source. This scenario is essentially 
different from the previous ones because Alice can discard not 
only the qubit which she could not control but also the other 
qubit from the pair. Bob never finds out that the pair ever 
existed. Alice can transfer comprehensible messages 
deterministically. 
\end{enumerate}

\begin{figure}[hbt]
\centerline{\psfig{file=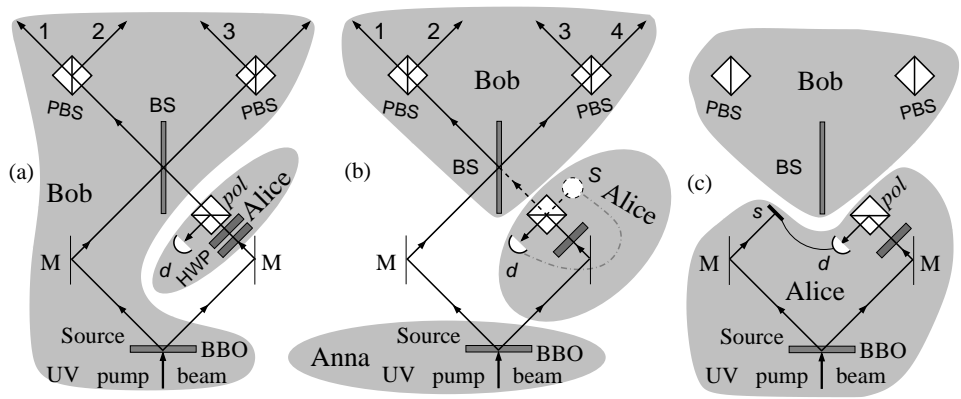,width=\textwidth}} 
\vspace*{8pt}
\caption{Superdense coding. Operational definitions (a), (b), 
and (c) that we consider in the text.}
\label{fig:matteo}
\end{figure}

These three operational scenarios are shown in Fig.~\ref{fig:matteo}.  
Interpretations of the scenarios essentially depend on applications. 
We elaborate on their applications in Sec.~\ref{sec:disc} and here 
we just discuss when the scenarios can be considered deterministic.  

\begin{enumerate}[(a)]
\item Alice is being sent her qubits and Bob expects her 
to use as many of them as she can. Alice sends $|\chi^1\rangle$
and $|\chi^2\rangle$ with an efficiency ideally approaching 
100\%. When sending $|\chi^3\rangle$ or $|\chi^4\rangle$
she has only 50\%\ probability of success, but she knows 
when she was successful and when not---her detector will 
not click when she was and will click when she was not. 
Bob will also know when Alice was not successful because 
he will then receive only one photon. So they can discard 
unsuccessful attempts. 
Now the question emerges whether we have an application 
for which it would be important to  worry about the lost 
photon pairs. If not, we can speak of ideally deterministic 
superdense coding. Application proposed in 
Sec.~\ref{sec:disc} supports it.
\item Anna is demanding and wants Alice to use all 
the photons all the photons she sends her. However, 
she expects of Alice only to sends states 
$|\chi^i\rangle$, $i=1,2,3,4$ as she can. So, when sending 
say $|\chi^3\rangle$ Alice sends half of them through her 
PBS as ``they choose'' and clone the other half with the help 
of quantum dots (deterministic cloning of definite known 
polarization is possible). We give an application of this
scenario in Sec.~\ref{sec:disc}.
\item Alice owns the source and both photons. 
She is allowed to manipulate just one photon but she 
can stop the other if her photon chooses a ``wrong''
exit from her PBS. 
Here the question emerges whether we can have any 
reason not to allow Alice to stop the whole pair. 
Again everything depends on the application. 
But in the absence of a dominant superdense coding 
application we can again speak of  ideally deterministic 
superdense coding. In Sec.~\ref{sec:disc} we give an 
application which makes use of such a coding. 
\end{enumerate}

\section{Discussion}
\label{sec:disc}

Efficient recognition of all four Bell basis states is 
undoubtedly essential for teleportation because they 
describe entanglement of photons which 
serve as ``carriers'' for teleportation. However, 
for superdense coding it is essential that we 
transfer four messages by manipulating just one 
of the originally entangled qubits. 

We showed that in the current absence of prevailing 
application of superdense coding we can carry it out 
deterministically with the linear optics in three different ways. 
Here we present some possible applications of the coding 
in quantum cryptography. As opposed to ``pure'' 
superdense coding, its cryptography application will
include classical channels but we keep the basic 
superdense coding scenarios from Sec.~\ref{sec:def}.
\begin{enumerate}[(a)]
\item Bob and Alice discard unsuccessful messages (33\%). 
Alice repeats every such message.
Information transfer with successful messages can be considered
deterministic in the absence of applications which would forbid 
discarding unsuccessful messages. Application can be the 
{\em ping-pong\/} quantum cryptography 
protocol.~\cite{bostrom-felbinger-02,bostrom-felbinger-08}  
Since in this protocol we do not have to have a classical channel 
through which Alice would inform Bob which messages to keep 
and which to discard as in BB84 protocol, Alice and Bob make a 
direct deterministic transfer of comprehensible messages with
their 67\%\ of messages. The transfer is done with four messages
per Alice's qubit and with linear optical elements. The discarded 
33\%\ of messages do not impair the quality of the transfer in 
any way. More over, in the ping-pong protocol they need not be
discarded but can be used as a control channel;
\item Alice takes care only to send all her photons as she can. 
So, she can send four different messages (four different photon 
states $|\chi^i\rangle$, $i=1,2,3,4$) by manipulating just one 
photon but does not have a control over half of the states she 
sends, although she deterministically knows which messages 
she sent. Application might again be a ping-pong protocol. 
Alice can inform Bob on the cloned photons (with a delay) over 
a public (classical) channel so that Bob can change the received 
``wrong'' message into the one Alice intended to send. The 
message is still unreadable to Eve provided Alice randomly 
changes the orientation of her qubit and informs Bob on it 
with a delay;  
\item Alice sends messages cleanly and deterministically to Bob 
by stopping both photons whenever her photons come from 
the ``wrong'' exit. Alice repeats every such message.
Bob does not know anything about the 
existence of the ``wrong'' pairs. In the ping-pong 
protocol Alice can use a public channel to tell Bob (with 
a delay) to erase his qubit from the pair containing 
Alice's ``wrong qubit.'' 
\end{enumerate}
In the end we would like to discuss the following possible 
objection to our approach: ``A superdense coding applies to 
message generation and not to a message recognition; therefore 
we cannot discard 25\%\ of ``wrong'' messages in our options 
(a) and (c); hence,  (a) and (c) are only an alternative 
scheme to achieve dense coding.'' The answer to this objection 
is simple. 

First, we do not discard 25\%\ but 33\%\ of messages
assuming that they are equally distributed. This is because for 
an  equal distribution of messages we have to compensate for
the messages ($|\chi^{3,4}\rangle$) that cannot be sent in 
half of Alice's attempts by increasing the number of her 
attempts to do so. Let $n$ be the the number of each of 
the four messages. Each unsuccessful attempt to send 
$|\chi^{3,4}\rangle$ Alice has to repeat 
repeat until the messages gets through.
That gives $4n+n+n=100\%$ and 
the percentage of each of the sent messages is 16.7\%. 
The percentage of each kind of ``wasted'' (repeated) messages 
is also 16.7\%\ and this reduces the efficiency 
of the four encoded messages by 33\%: $4\cdot 0.67=2.7$
Hence, the channel capacity of the dense coding with respect 
to the total amount of the photon pairs generated at the source 
($\log_23=1.585$)  is higher then the channel capacity of our 
protocol ($\log_22.7=1.433$) but our protocol transfers 4 
messages, while the dense coding transfers only three. 

Second, we deterministically generate four different 
messages {\em after\/} first discarding 33\%\ of unusable 
detections. Hence, our protocol is not an alternative scheme
of dense coding.  In particular, 
\begin{enumerate}
\item pair generation of photon pairs in the photon source 
and message generation are two independent things; there is 
no physical reason why Alice and Bob should not be  
allowed to shrink the number of photons they obtain 
from the crystal, for Alice's generation of  messages; 
\item both, our protocol and the dense coding protocol are 
about message generation but with our protocol we are able 
to transfer four messages, while with the dense coding we 
can transfer only three; in dense coding Bob cannot discriminate 
between two of four messages while in our protocol he can 
deterministically discriminate all four ($|\chi^i\rangle$, 
$i=1,2,3,4$) messages. 
\item if any of the two protocol can be considered
non-deterministic it is the dense coding one because 
there Bob cannot discriminate between $
|\Phi^-\rangle$ and $|\Phi^+\rangle$. 
\end{enumerate}
As for point (1) above we stress that similar 
discarding of unwanted events is a standard technique 
of the quantum information engineering. Consider, e.g.,
generation of entangled photons on demand from 
three spontaneous parametric down-conversion 
sources.~\cite{wagenkecht-pan-10,barz-zeilin-10,kok-10}
We discard photon detections after photon detections until 
we finally get a right set of four detections that tell us that
the remaining two photons are entangled and ready for 
usage. Actually we discard so many of them that within
a required time window we have a succes probability of 
the order of $10^{-6}$. In this procedure a detection of 
four photons determines the entangled photons on 
demand and in our procedure Alice's manipulation 
of her qubits determines the number of superdense 
coded pairs; discarded pairs are irrelevant for the coding 
and play no role in it; relevant are only those that 
carry Alice's messages to Bob. 

Therefore, the claim ``A superdense coding 
applies to message generation and not to a message 
recognition'' seems to be a matter of taste. (See ArXiv 
v.~1 \&\ 2.)

\section*{Acknowledgments}
Supported by the Ministry of Science, Education, and Sport
Sport of Croatia through project No.~082-0982562-3160.

\bibliographystyle{plain}

\end{document}